\documentclass[10pt, conference, compsocconf]{IEEEtran}
\usepackage{cite}

\usepackage{graphicx}
\usepackage{hyperref}
\usepackage{breakurl}

\ifCLASSINFOpdf
\else
\fi
\usepackage{tikz}
\usepackage{textcomp}
\usepackage{hyperref}
\usepackage{lipsum}

\newcommand\copyrighttext{%
  \scriptsize \textcopyright 2020 IEEE. Personal use of this material is permitted. Permission from IEEE must be obtained for all other uses, in any current or future media, including reprinting/republishing this material for advertising or promotional purposes, creating new collective works, for resale or redistribution to servers or lists, or reuse of any copyrighted component of this work in other works. doi: {10.1109/AIVR50618.2020.00083}}
\newcommand\copyrightnotice{%
\begin{tikzpicture}[remember picture,overlay]
\node[anchor=south,yshift=10pt] at (current page.south) {\fbox{\parbox{\dimexpr\textwidth-\fboxsep-\fboxrule\relax}{\copyrighttext}}};
\end{tikzpicture}%
}

\begin{document}
%
\title{Eye Tracking Data Collection Protocol for VR for Remotely Located Subjects using Blockchain and Smart Contracts}


\author{\IEEEauthorblockN{Efe Bozkir,
Shahram Eivazi, Mete Akg{\"u}n and
Enkelejda Kasneci}
\IEEEauthorblockA{Department of Computer Science,
University of T{\"u}bingen\\
T{\"u}bingen, Germany\\
Email: \{efe.bozkir, shahram.eivazi, mete.akguen, enkelejda.kasneci\}@uni-tuebingen.de}}


%


\maketitle
\copyrightnotice

\begin{abstract}
Eye tracking data collection in the virtual reality context is typically carried out in laboratory settings, which usually limits the number of participants or consumes at least several months of research time. In addition, under laboratory settings, subjects may not behave naturally due to being recorded in an uncomfortable environment. In this work, we propose a proof-of-concept eye tracking data collection protocol and its implementation to collect eye tracking data from remotely located subjects, particularly for virtual reality using Ethereum blockchain and smart contracts. With the proposed protocol, data collectors can collect high quality eye tracking data from a large number of human subjects with heterogeneous socio-demographic characteristics. The quality and the amount of data can be helpful for various tasks in data-driven human-computer interaction and artificial intelligence.
\end{abstract}


%
\IEEEpeerreviewmaketitle

\section{Introduction}
Over past decades, head-mounted display (HMD) technologies have taken advantage of innovations from imaging and eye tracking research to improve image quality and utility of user interfaces. To date, several consumer level HMDs have integrated eye trackers, providing opportunity for researchers to collect eye movement data for user behavior analysis and data-driven interaction.

In the virtual reality (VR) context, it has been shown that eye tracking is helpful for assessing human attention~\cite{Bozkir:2019:ADA:3343036.3343138}, detecting human stress~\cite{stress_vr}, assessing cognitive load~\cite{8797758}, predicting human future gaze locations~\cite{8998375}, supporting evaluation and diagnosis of diseases~\cite{7829437}, motion sickness detection~\cite{8642906}, foveated rendering~\cite{Arabadzhiyska2017, 9005240}, continuous authentication~\cite{Zhang:2018:CAU:3178157.3161410}, gaze-based interaction~\cite{VRPursuits_interaction}, training~\cite{8448290}, and redirected walking~\cite{redirected_walking_steinicke}. Many of these tasks are data-driven and require a large quantity of eye tracking data which are usually collected in laboratory settings. Subjects are frequently compensated with some amount of money or gifts for their participation. Two drawbacks of these settings are the lack of heterogeneity in socio-demographic characteristics of data collected subjects and potential for unnatural behaviors of subjects due to the constraints of the laboratory settings. While VR is a unique and controlled environment and requires dedicated hardware such as HMDs, as personal usage of such devices increases, we foresee that it should be possible to collect data from remotely located subjects, i.e., at their homes. Especially in situations such as COVID-19, this possibility could help experimental works continue in a remote setting. Currently, for crowd-sourcing or similar purposes, platforms such as Amazon Mechanical Turk\footnote{\url{https://www.mturk.com/}} are used. While it is not possible to collect VR data with such platforms, for other types of data collection significant compensations are paid to manage the remote subjects' work. In addition, these third-party platforms store and manage data. In fact, as eye tracking and movement data represent unique information about the subjects, the data manipulation possibility of the third parties should be prevented. Third parties should only act as a bridge between the data collector and the subjects in case there is no direct communication between the parties.

To overcome the disadvantages of the laboratory setting and enable remotely located subject participation in eye tracking experiments in the VR context, we propose a blockchain-based protocol on the Ethereum blockchain using smart contracts, where we use the blockchain for validation of data integrity and smart contract for compensation management. For this study, we focus mainly on collecting eye tracking data in VR environments as many modern HMDs come with integrated eye trackers. This means that subjects do not need any additional effort to integrate any sensor into their setup. It is relatively easier to control environmental configurations in HMDs when compared to other setups such as illumination and light-sources which may affect subject behaviors or eye movement patterns. However, the proposed protocol can also be used in similar setups as long as identical experiment configurations are guaranteed.

While the first prominent usage of the blockchains is Bitcoin~\cite{bitcoin_whitepaper} and most of the applications are in the financial domain, blockchains also draw attention of the human-computer interaction (HCI), eye tracking, and VR communities. Opportunities and challenges for the HCI and interaction design and the role of HCI community were discussed in ~\cite{foth_blockchain_for_interaction_design} and ~\cite{making_sense_blockchain_hci}, respectively. An augmented reality (AR)-based cryptocurrency wallet was developed in~\cite{crypto_ar_wallet} to familiarize users with blockchain wallet services. In addition, GazeCoin is a cryptocurrency for VR/AR which is exchanged between content makers, advertisers, and the users~\cite{gaze_coin_white_paper}. Apart from the financial use-cases, due to their immutability blockchains are used as notary. Additionally, Ethereum platform brings the smart contract~\cite{Szabo_1997} concept to the blockchains~\cite{eth_white_paper}. One of the straightforward usages of smart contracts is escrow services. For the remote purchase of goods, buyer and seller parties use the smart contracts without trusting one another and a trusted centralized party during the escrow. The smart contracts that are deployed on the blockchains distribute the money to the parties once buyer and seller parties fulfill their obligations in the remote purchase. In our protocol, we treat recorded eye tracking data as digital good so that compensation distribution is done by the smart contracts. To assure that the recorded data are not altered by the subjects, the hash of the recorded data using white-box cryptography~\cite{whitebox_crypto_alex_biryukov} is stored in the blockchain, which enables the blockchain as a notary for data integrity. Our major contributions are as follows. 
\begin{itemize}
    \item A blockchain-based eye tracking data collection protocol for remotely located subjects that can be used for eye tracking experiments in VR, which presents the opportunity to collect data from a various number of subjects.
    \item Delegation of mutual trust issues for compensation management and integrity of the recorded eye tracking data to smart contracts and blockchains, respectively.
    \item Elimination of the centralized third parties for compensation management, data collection and manipulation, which is optimal from a privacy perspective.
\end{itemize}

\section{Preliminary Definitions}
As our protocol consists of interdisciplinary work from different domains such as virtual reality, blockchains, and cryptography, we provide some definitions that are used throughout the paper.

\textbf{Blockchain~\cite{bitcoin_whitepaper}:} An immutable ledger that consists of a chain of blocks that keeps records of transactions, maintained by several machines in a peer-to-peer network. Each block consists of a timestamp, transaction data, and the cryptographic hash of the previous block. As each block consists of the cryptographic hash of the one prior, immutability is automatically preserved unless one party has the majority of the computational power.

\textbf{Ethereum~\cite{eth_white_paper}:} Public, open-source, blockchain-based, and smart contract supporting distributed platform. 

\textbf{Ether (ETH)~\cite{eth_white_paper}:} The cryptocurrency of the Ethereum platform.

\textbf{Smart Contract~\cite{eth_white_paper}:} A self-executing, irreversible, and transparent contract between buyer and seller, implemented in the code.

\textbf{White-box cryptography~\cite{wyseur_whitebox}:} ``Software protection technology which allows for the application of cryptographic operations without revealing any critical information such as secret keys.''

\section{Protocol}
In this section, we discuss our protocol and its flow, assumptions, and details of the implementation.

\subsection{Flow}
\label{subsection:flow}
Our proposed protocol consists of two parties as data collector and subjects. The data collector is responsible for providing the VR application for eye tracking data collection and subjects are tasked with carrying out the experiment and providing the recorded eye tracking data. At the end of a valid experiment, subjects are compensated for their participation. Let us assume that each subject is compensated with $X$ unit of ETH for the valid data recorded from an experiment session. A relevant amount can be set for compensation depending on the experiment.

\begin{figure*}[!ht]
  \centering
   \includegraphics[width=\linewidth]{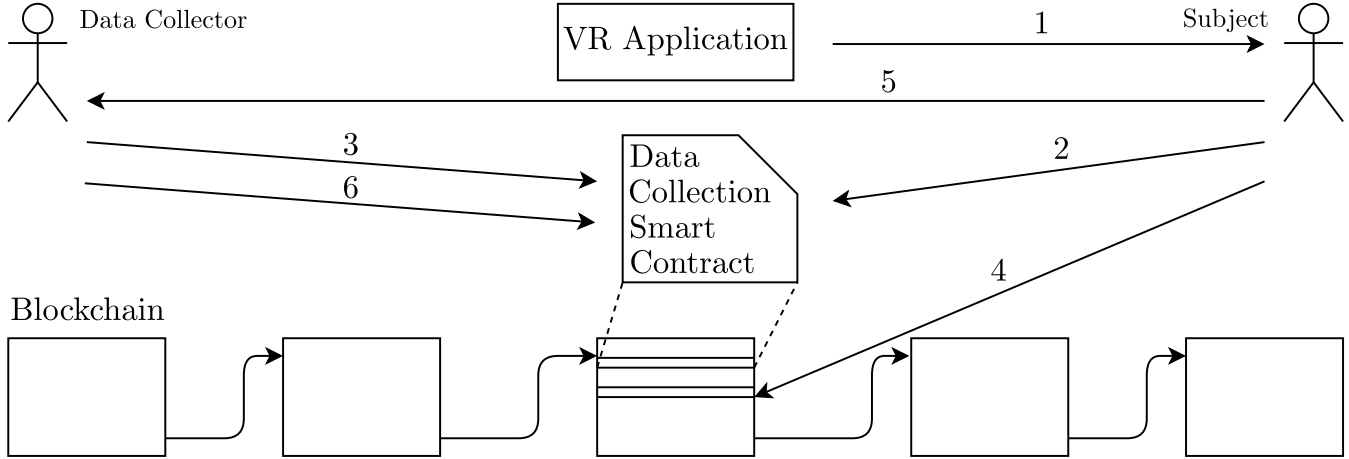}
  \caption{Blockchain-based protocol and its steps. ($1$) Subject fetches the application and carries out the experiment. ($2$) Subject initiates the smart contract. ($3$) Data collector confirms the contract creation and stakes. ($4$) Subject stores the recorded data hash in blockchain. ($5$) Subject transfers the recorded data to the data collector. ($6$) Data collector confirms the data collection.}
  \label{fig:protocol}%
\end{figure*}

Figure~\ref{fig:protocol} shows the overall flow and short descriptions of each step of the protocol. As the \textbf{step 1}, subjects fetch the application from the data collector and carry out the experiment. While the content of the stimuli changes depending on the use-case, the VR application validates the eye tracking data quality at the end of each experimental session by using tracking rates or confidence intervals that are provided by the eye tracker. If the recorded eye tracking data are too noisy, subjects are not supposed to send the data to the data collector, where they are informed by the VR application. This obligation forces the subjects to follow the instructions of the VR experiment, such as eye tracker calibration, carefully while the data collector obtains better quality data in the end. After the validation success, the VR application calculates the hash output of the recorded eye tracking data and saves it. Saving the hash output is required for assessing the data integrity; however, adversarial subjects can easily find out the hashing algorithm using the executable of the VR application on their own devices. Therefore, we opt for a white-box~\cite{wyseur_whitebox,whitebox_crypto_alex_biryukov} paradigm for calculating the data hash. In the white-box paradigm, the adversary is supposed to have visibility of the inputs, outputs, and other intermediate steps. White-box cryptography achieves protection of confidential information such as secret keys while keeping the application semantically the same. Even if adversaries infer the hash function, due to the lack of secret key, it is not possible to generate a hash output for altered data. Consequently, subjects are obliged to behave honestly, where honest behavior means not altering the recorded data. In the end of the first step, once the recorded data is validated and hash value is saved, the subjects are informed by the VR application that the recorded eye tracking data is reportable.

As the \textbf{step 2}, the subjects initiate the smart contract and stake double the amount of compensation, which is $2X$ ETH for our case. Staking double the amount of compensation that they will obtain from the smart contract forces subjects to act honestly; otherwise, they lose the amount that they stake. As the \textbf{step 3}, the data collector confirms the data collection and stakes the same amount as the subject, which is $2X$ ETH to the smart contract. While the compensation is $X$ ETH per experiment, the data collector is supposed to stake double the amount of compensation so that it also becomes an obligation to behave honestly. Otherwise, the doubled amount of compensation will be lost without obtaining the recorded data. As the \textbf{step 4}, the subjects store the hash output that is reported by the VR application in the blockchain and, as the \textbf{step 5}, they send the recorded data along with the transaction hash of the transaction for storing the data hash in the blockchain to the data collector. If subjects try to alter the data, the hash in the blockchain and the altered data will not match and it will be discovered by the data collector. As the \textbf{step 6}, the data collector obtains the recorded eye tracking data and transaction hash of the data hash and checks whether or not the obtained data and the hash provided by the subjects overlap using the hash function that is implemented in the VR application and secret keys. If the reported data and hash value stored in the blockchain overlap, the data collector confirms the smart contract and that the obtained data are valid. Then, the smart contract automatically distributes $3X$ and $X$ ETH to the subject and the data collector, respectively. In the end, each subject earns $X$ unit of ETH for participation in the experiment, where the data collector obtains the recorded eye tracking data. Due to the immutable nature of blockchains and smart contracts, none of the parties can alter the values in the blockchain and behave as an adversary.

In the protocol, as both parties stake more than the amount they are supposed to spend or earn, they have to act honestly in order to achieve successful data collection and compensation distribution, otherwise data collection is not finalized and parties lose the amount they stake. In particular, the subjects have to stake double the amount of compensation that they will receive whereas the data collectors have to stake double the amount of compensation that they will give. Since the smart contracts are immutable and stored in the blockchain, a third-party application is not needed for compensation distribution or data manipulation, which is useful from a privacy preservation point of view.

\subsection{Assumptions}
We have three main assumptions in our protocol. Firstly, validation of the quality of the recorded eye tracking data is automatically completed by the VR application at the end of each experiment by using metrics such as tracking ratio or confidence levels reported by the eye tracker. Due to poor calibration for eye tracking, removal of the head-mounted display (HMD) in the middle of experiment, or similar reasons, recorded eye tracking data may have an extensive amount of noise level. Instead of cleaning data offline extensively after the experiments, our protocol assumes that data validity is checked at the end of each experiment by the VR application and the application informs the subjects whether the quality of the data is valid and reportable.

Secondly, the recorded eye tracking data is hashed using white-box cryptography and stored at the end of the experiment by the VR application to be stored in the blockchain for validation of the data integrity. In traditional eye tracking experiments, subjects participate in the experiments on the devices that are provided by the data collectors. However, in the remotely located subject participation, subjects run the applications on their own devices. Therefore, they have direct access to the provided application and if any adversarial subject analyzes the binary implementation of an application that does not use white-box paradigm, they can easily infer the used hash function and generate hash output for fake data. On the contrary, when using white-box cryptography, the secret keys are not leaked even if adversaries analyze the binary implementation. Even if an adversary infers the hash function, a hash output for fake data cannot be generated without secret keys. Therefore, white-box paradigm is used by the VR application. If subjects alter the recorded data or send fake data to the data collector, the generated hash value will not match the recorded data, which leads subjects to lose their staked compensation in the smart contract.

Lastly, as our protocol does not use any centralized third party, a secure direct communication is needed for exchanging the application and the recorded data between the data collector and subjects. In case it is not available, a bridging third party only for communication purposes can be implemented.

\subsection{Implementation}
We select the Ethereum platform for our proof-of-concept due to its public blockchain, relatively higher number of nodes, and status as one of the most mature platforms in the blockchain domain. However, any blockchain-based platform that supports smart contracts can be opted in.

We implement the blockchain related part of the protocol, particularly the steps $2$, $3$, $4$, and $6$ discussed in Section~\ref{subsection:flow}, using Solidity\footnote{\url{https://docs.soliditylang.org/}} and a simple purchase smart contract~\cite{ng_smart_contract} on the Ropsten Testnet of the Ethereum platform. In the beginning of the data collection, both the data collector and the subject hold $1$ ETH in their wallets. We select the compensation amount as $0.025$ ETH. For the calculation of the hash output of the recorded eye tracking data, we use synthetic data; however, any eye tracker integrated to modern HMDs can be used in a real-world implementation. The hash value of the data is calculated using Keyed-Hashing for Message Authentication (HMAC)~\cite{hmac_97} and Secure Hash Algorithm$3$-$512$ (SHA$3$-$512$)~\cite{sha3_standard} as it is possible to have white-box implementation of the HMAC. The calculated hash value is stored in the input data field of a self transaction from the subject. After the protocol execution, the data collector and the subject hold $\approx 0.975$ and $\approx 1.025$ ETH when the transaction fees are subtracted, respectively. The smart contract, overall procedure, the data collector, and the subject parties are available on the Ropsten Testnet via following link: \url{https://ropsten.etherscan.io/address/0x0e937a4a4618dd8d5a12ec4a9f8fd61d6bfd13e4}.

In the above link, there are three transactions in chronological order that correspond to steps $2$, $3$, and $6$ of our protocol. The subject (address starting with $0x89$) and the data collector (address starting with $0x44$) of our implementation are available in the source of the first and the second transactions of the smart contract, respectively. There are three transactions in the subject address. The first and second transactions are for depositing the test ETH and initiating the smart contract, respectively. The third transaction in the subject address is a self transaction and corresponds to step $4$ of our protocol. In the ``Input Data'' field of the self-transaction, the calculated data hash is available.

\section{Conclusion and Discussion}
We proposed a blockchain-based protocol for collecting eye tracking data in VR from remotely located subjects. As eye tracking experiments are usually conducted in laboratory settings with a limited number of subjects from similar backgrounds in terms of socio-demographic characteristics, it is a challenge to draw generic data-driven conclusions. Due to the laboratory settings, subjects may not behave naturally. While our protocol overcomes the drawbacks of the traditional eye tracking data collection setups without needing a centralized third party for data collection and compensation management, it also creates an opportunity to carry out the data collection anonymously, which is optimal for the privacy of subjects. We focused on the eye tracking data collection in VR setups as validation of the eye tracking data and generation of the controlled environments with VR can be done easily. In addition, current availability of eye tracker integrated HMDs in the consumer market supports our protocol for VR and eye tracking data; however, the proposed protocol may be useful for other types of eye trackers, sensors, or environments as long as identical configurations between subjects can be generated. In contrast to traditional eye tracking experiments, subject consent, additional questionnaire, or similar information should be collected digitally using our protocol. Our protocol may also require an application-level effort to have one-to-one mapping between subjects and experiments.

As future work, we plan to have an end-to-end implementation of our protocol along with a real VR application and HMD-integrated eye tracker. In addition, while transactions are applied anonymously on the public blockchains, it is possible to track them. Recent work on eye tracking, HCI, and VR~\cite{bozkir_ppge, steil_diff_privacy, fuhl2020reinforcement, sumer2020automated, bozkir2020differential} emphasize the importance of privacy preservation. Combining privacy-preserving methods with our protocol remains as part of future work.
\vspace{-0.1em}
\section*{Acknowledgments}
E.B. thanks Batuhan Sar{\i}o\u{g}lu for useful discussions on blockchains.
\vspace{-0.2em}
\bibliographystyle{IEEEtran}
\bibliography{IEEEabrv,refs}
%

\end{document}